\documentclass[aps,prb,superscriptaddress,twocolumn,showpacs]{revtex4}
\usepackage{amsmath}
\usepackage{graphicx,epsfig}
\usepackage{bm}

\renewcommand{\v}[1]{{\bf #1}}
\newcommand{\be}{\begin{equation}}
\newcommand{\ee}{\end{equation}}
\newcommand{\bd}{\begin{displaymath}}
\newcommand{\ed}{\end{displaymath}}
\newcommand{\ba}{\begin{eqnarray}}
\newcommand{\ea}{\end{eqnarray}}
\newcommand{\nn}{\nonumber \\}

\newcommand{\bpm}{\begin{pmatrix}}
\newcommand{\epm}{\end{pmatrix}}

\begin{document}

\title{Zero-temperature phases for chiral magnets in three dimensions}

\author{Jin-Hong Park}
\affiliation{Department of Physics and BK21 Physics Research
Division, Sungkyunkwan University, Suwon 440-746, Korea}
\author{Jung Hoon Han}
\affiliation{Department of Physics and BK21 Physics Research
Division, Sungkyunkwan University, Suwon 440-746, Korea}

\date{\today}

\begin{abstract} We calculate the energies of
various multiple-spiral spin configurations for the three-dimensional
model of chiral magnets. The ground-state phase diagram is obtained
in the space of anisotropy, magnetic field, and interaction
parameters. Regimes with multiple-spiral, or spin crystal ground
states of two- and three-dimensional nature are identified. The
Skyrmion number responsible for anomalous Hall effect is examined for
each spin crystal configuration. A geometric interpretation is given
for the three-dimensional spin crystal lattice structures as the
periodic arrangement of hedgehog and anti-hedgehog singularities.
\end{abstract}
\maketitle

\section{Introduction}
Skyrmionic spin texture is a natural excitation of a ferromagnet in
two spatial dimensions (2D) carrying nontrivial topological numbers
\cite{skyrme,raj}. Although their creation energy remains finite (of
order of several $J$, $J$=ferromagnetic exchange energy) and can be
excited thermally, an important work of
Polaykov\cite{polyakov,nagaosa} demonstrated that non-topological
spin-wave excitations already destroy order at all finite
temperatures, pre-emptying the chance of a phase transition driven
by topological defects. This is in contrast to the XY spin model in
the same dimension, where the transition to disordered phase indeed
occurs by proliferation of topological defects known as
vortices\cite{KT}.

Recent advances in the experimental investigation of chiral
magnets\cite{pfleiderer-1,pfleiderer-2,pfleiderer-3,tokura,AHE-1,AHE-2}
brought new insights into the role and importance of Skyrmions in
the class of magnets characterized by both Heisenberg ($J$) and
Dzyaloshinskii-Moriya ($D$) interactions. A most interesting finding
of the recent experiments is that a rather small magnetic field of
order $D^2/J$ leads to the proliferation of Skyrmions and their
crystallization into a columnar
(3D)\cite{bogdanov-1,bogdanov-2,pfleiderer-1} or a pancake-like
(2D)\cite{tokura}  triangular array, depending on the
dimensionality. The stability of the triangular Skyrmion lattice was
successfully demonstrated based on Ginzburg-Landau
theory\cite{pfleiderer-1} and on microscopic model
calculations\cite{YONH,han}.

Much less understood aspect of Skyrmion physics in chiral magnets is
whether it is possible to realize other forms of Skyrme crystalline
order. Indeed it was argued that a square array of Skyrmions and
anti-Skyrmions would be possible
energetically\cite{YONH,bogdanov-nature,bogdanov-arxiv}. In three
dimensions, study of different 3D crystalline structures was
pioneered by Binz and collaborators\cite{binz}. In this paper, we
carry out a comprehensive study of the energetic stability of many
different kinds of spin crystal configurations in three spatial
dimensions, under the influence of a number of spin anisotropy
interactions and magnetic field. Our calculation focuses on
zero-temperature energetics and differs from other considerations
where temperature control also played a vital
part\cite{pfleiderer-1,bogdanov-nature,bogdanov-arxiv}. It also
differs from Binz \textit{et al.}\cite{binz} in that the role of
magnetic field in stabilizing Skyrmion order is more thoroughly
examined here.

Some conclusions reached in our paper are as follows. (i)
Two-dimensional multiple-spiral structures are energetically
unfavorable in three spatial dimensions. (ii) For strong
interactions fully three-dimensional spin crystal phase becomes more
stable compared to single-spiral phases. Typically, magnetic field
tends to stabilize 3D spin crystal structures. (iii) All 3D
multiple-spiral structures can be understood as a lattice of
hedgehogs and anti-hedgehogs. Such identification allows a
complicated, but still straightforward evaluation of the Skyrmion
number. Organization of the paper begins by introducing several
multiple spiral states in Sec. \ref{sec:model} along with some
single spiral states. The free energy for each state is worked out
following the detailed procedure described in the Appendix. In Sec.
\ref{sec:phase-diagram} the ground state phase diagram is worked out
in search for the energetic stability of multiple spiral phases in
three dimensions. We also carry out a calculation of the spin
chirality (Skyrmion number) for each of the multiple spiral phases
in Sec. \ref{sec:chirality}. A useful visual interpretation is given
of the three-dimensional Skyrmion crystals as a lattice of hedgehog
singularities. Finally we conclude with a summary in Sec.
\ref{sec:summary}.

Throughout the paper, the expression ``multiple spiral" state is
used interchangeably with ``spin crystal" state. A single spiral
state does not form a spin crystal.

\section{Multiple Spiral State Model}
\label{sec:model}

The multiple spiral spin configuration can be written in the general
form

\ba \v n_{\v r} &=& \v n_{\v 0} + \v \Phi_{\v r}, \nn
\v \Phi_{\v r} &=& \sum_{\alpha=1}^N \Bigl( \v n_{\alpha} e^{i \v
k_\alpha \cdot \v r } +\v n_{\alpha}^* e^{-i \v k_\alpha \cdot \v r }
\Bigr),\label{eq:n-alpha} \ea
as the sum of a uniform magnetization $\v n_0 = (0, 0, n_0)$ due to
the Zeeman field and $N$ independent spirals of wave vectors $\v
k_\alpha$, $\alpha =1 \cdots N$. It is assumed that all $\v
k_\alpha$'s are of the same length, and the amplitudes $\v n_\alpha
\cdot \v n_\alpha^* = n_H^2 /2$ are the same for all the spiral
modes. The magnitude of the spin is position-dependent, and we
impose the average normalization on $\v n_{\v r}$

\ba \langle{\v n_{\v r}^2}\rangle = 1= \v n_0^2 + 2 \sum_\alpha \v
n_\alpha \cdot \v n_\alpha^* =n_0^2 + N n_H^2 . \ea
In this sense, our spins are ``soft", with the origin of softness
possibly traced to the coupling of spins to the conduction electron
bath or the averaging over fast fluctuations.

For right-handed spirals we can choose

\ba \v n_\alpha &=& {n_H \over 2}e^{i\theta_\alpha}
(\hat{e}_{1\alpha}-i \hat{e}_{2\alpha}),\ea
using a pair of orthonormal vectors forming a triad,
$\hat{e}_{1\alpha} \times \hat{e}_{2\alpha} = \hat{k}_\alpha$. An
arbitrary phase $\theta_\alpha$ is assigned for each spiral mode for
generality.

Our consideration of the energy of multiple spiral spins is based on
the free energy

\ba F[\v n ] &=& \int d^3 \v r ~ f [\v n ] , \nn
f[\v n ] &=& {1 \over 2} A(n_x^2 + n_y^2 + n_z^2) + D \v n \cdot
(\bm \nabla \times \v n) -\v B \cdot \v n \nn
&+& {1 \over 2}J
[(\bm \nabla n_x)^2+(\bm \nabla n_y)^2+(\bm \nabla n_z)^2] \nn
&+& {1 \over 2} C \left[\Bigl({\partial n_x \over \partial
x}\Bigr)^2 + \Bigl({\partial n_y \over \partial y}\Bigr)^2 +
\Bigl({\partial n_z \over \partial z}\Bigr)^2 \right] \nn
&+& U \left( n_x^2 + n_y^2 + n_z^2 \right)^2 + W \left(n_x^4 + n_y^4
+ n_z^4 \right) . \label{eq:free-energy}\ea
Ferromagnetic exchange and Dzyaloshinskii-Moriya (DM) interactions
are given by $J$ and $D$ while $C$, $U$, and $W$ are the spin
anisotropy and interaction terms up to fourth order introduced by Bak
and Jensen\cite{bak}. Zeeman field $\v B$ is always directed along
the $\hat{z}$-direction in this paper. The $A$-term can be dropped in
practice because we will always require average magnetization of
unity, which also makes our calculation effectively zero-temperature.
The integration $\int d^3 \v r$ extends over three-dimensional space
and is divided by the space volume.

If we remove all but  $J$, $D$, and $\v B$ from the free energy, the
multiple spiral state defined in Eq. (\ref{eq:n-alpha}) can be
readily shown to have the energy

\ba  E  = {1 \over 2}N n_H^2 \Bigl(J  k^2 -2 D  k \Bigr)- B n_0 .
\ea
Upon optimizing $k$ at $k= D/J$ it becomes

\ba E = - {1 \over 2} {D^2\over J}N n_H^2 - B n_0 = {1 \over 2}
{D^2\over J}(n_0^2 -1) - B n_0. \ea
Minimizing it further in terms of $n_0$ gives

\ba n_0= {B \over D^2/J}, ~~ E= -{1\over 2} {B^2 \over D^2/J} .
\label{eq:quadratic-E}\ea
It is interesting that all multiple spiral states of the type
(\ref{eq:n-alpha}), regardless of the number of spiral components
$N$ and their orientations, carry exactly the same $\v k$-vector
length, magnetization $n_0$, and the ground state energy, in the
presence of magnetic field.

It is then up to the anisotropy and interaction terms $C,U,W$ to
separate out the energies of different multiple spiral states. Below
we work with eight exemplary spin configurations, drawn from both
recent experiments and other theories, with $N$, the number of
spiral components, ranging from 1 to 6.  For $N>1$ one will obtain
spin crystal state of some sort.

First, three kinds of $N=1$ single spiral states are considered.

\begin{itemize}

\item Helical spin state with $\v k =
(k/\sqrt{3})(1,1,1)$ has  the spin configuration

\ba \v n^\mathrm{H_{111}}_{\v r} &=& \v n_0 + \v n_1^\mathrm{H_{111}}
e^{i \v k \cdot \v r} + (\v n_1^\mathrm{H_{111}})^* e^{-i \v k \cdot
\v r}, \nn
\v n_1^\mathrm{H_{111}} &=& {n_H \over 2}e^{i\theta} \left({\hat{x}
+\hat{y}- 2\hat{z} \over \sqrt{6}} + i{ {\hat{x} - \hat{y}} \over
\sqrt{2}}\right). \ea
This state appears as the ground state at low magnetic field
($B<B_c$) for bulk MnSi
crystal\cite{pfleiderer-1,pfleiderer-2,pfleiderer-3}.

\item  Another helical state has $\v k=
(k/\sqrt{2}) (1, 1, 0)$ and the spin configuration

\ba \v n^\mathrm{H_{110}}_{\v r} &=& \v n_0 + \v n_1^\mathrm{H_{110}}
e^{i \v k \cdot \v r} + (\v n_1^\mathrm{H_{110}})^* e^{-i \v k \cdot
\v r}, \nn
\v n_1^\mathrm{H_{110}} &=& {n_H \over 2}e^{i\theta} \left(\hat{z} +i
{{\hat{y}-\hat{x}} \over \sqrt{2}} \right). \ea
This state appears as the low-field ground state in the 2D
geometry\cite{tokura,YONH,han}.

\item Conical state has
$\v k = k(0, 0, 1)$ and the spin configuration

\ba \v n^\mathrm{Co}_{\v r} &=& \v n_0 + \v n_1^\mathrm{Co} e^{i \v k
\cdot \v r} + (\v n_1^\mathrm{Co} )^* e^{-i \v k \cdot \v r}\nn
\v n_1^\mathrm{Co} &=& {n_H \over 2}e^{i\theta} (\hat{x} -
i\hat{y}). \ea
It includes the ferromagnetic (FM) state as a special case. This is
the ground state for 3D MnSi at fields $B>
B_c$\cite{pfleiderer-1,pfleiderer-2,pfleiderer-3}.

\end{itemize}

With $N=2$ and $N=3$ we consider two kinds of two-dimensional Skyrme
crystal states extended along the $z$-direction in a columnar
manner.

\begin{itemize}

\item Square-lattice Skyrmion crystal (SkX2) $(N=2)$ has the propagation
vectors $\v k_1 = k(1, 0, 0)$, $\v k_2 = k(0, 1, 0)$, and the spin
configuration

\ba  \v n^\mathrm{SkX2}_{\v r} &=& \v n_0 + \sum_{\alpha} \Bigl( \v
n_\alpha^\mathrm{SkX2} e^{i \v k_\alpha \cdot \v r} + (\v
 n_\alpha^\mathrm{SkX2} )^* e^{-i \v k_\alpha \cdot \v r} \Bigr) ,\nn
 \v n_1^\mathrm{SkX2} &=& {n_H \over 2}e^{i\theta_1} (\hat{y} -
i\hat{z}),\nn
\v n_2^\mathrm{SkX2} &=& {n_H \over 2}e^{i\theta_2} (\hat{z} -
i\hat{x}). \ea
The spin configuration is that of half-Skyrmions and
half-anti-Skyrmions alternating on a square lattice. Such a state
was previously considered in Refs.
\onlinecite{bogdanov-nature,YONH,bogdanov-arxiv}. Its existence in
nature remains elusive so far.

\item  Triangular-lattice SkX state $(N=3)$ has $\v k_1 =k(1,0,0)$,
$\v k_2 =k(-1/2,\sqrt{3}/2,0)$, $\v k_3 =k(-1/2,-\sqrt{3}/2,0)$, and
the spin configuration

\ba \v n^\mathrm{SkX3}_{\v r} &=& \v n_0 + \sum_\alpha \Bigl( \v
n_\alpha^\mathrm{SkX3} e^{i \v k_\alpha \cdot \v r} + ( \v
n_\alpha^\mathrm{SkX3})^* e^{-i \v k_\alpha \cdot \v r}  \Bigr),\nn
\v n_1^\mathrm{SkX3} &=& {n_H \over 2}e^{i\theta_1} \left(\hat{z} +
i\hat{y}\right),\nn
\v n_2^\mathrm{SkX3} &=& {n_H \over 2}e^{i\theta_2}\left(\hat{z} -
i{\sqrt{3} \over 2}\hat{x}- i{\hat{y}\over 2}\right),\nn
\v n_3^\mathrm{SkX3} &=& {n_H \over 2}e^{i\theta_3}\left(\hat{z} +
i{\sqrt{3} \over 2}\hat{x} - i{\hat{y}\over 2}\right). \ea
This state was found to exist in the A-phase of MnSi and related
B20-compounds\cite{pfleiderer-1,pfleiderer-2,pfleiderer-3} as well
as in the 2D thin-film sample at low temperature in the presence of
magnetic field\cite{tokura}.

\end{itemize}
Another possible spin crystal arrangement with $N=3$ is the
simple-cubic (SC) multiple-spiral spin structure\cite{binz} which
consists of $\v k_1 = k(1, 0, 0)$, $\v k_2 = k(0, 1, 0)$ and $\v k_3
= k(0, 0, 1)$.

\ba  \v n^\mathrm{SC}_{\v r} &=& \v n_0 + \sum_{\alpha} \Bigl( \v
n_\alpha^\mathrm{SC} e^{i \v k_\alpha \cdot \v r} + ( \v
n_\alpha^\mathrm{SC})^* e^{-i \v k_\alpha \cdot \v r} \Bigr),\nn
\v n_1^\mathrm{SC} &=& {n_H \over 2}e^{i\theta_1} (\hat{y} -
i\hat{z}),\nn
\v n_2^\mathrm{SC} &=& {n_H \over 2}e^{i\theta_2} (\hat{z} -
i\hat{x}), \nn
\v n_3^\mathrm{SC} &=& {n_H \over 2}e^{i\theta_3} (\hat{x} -
i\hat{y}). \ea

\begin{figure}[ht]
\includegraphics[scale=0.4]{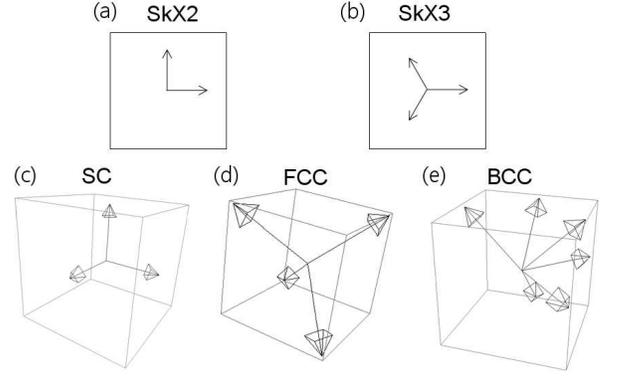}
\caption{The $\v k$-vectors of five multiple-spiral spin states
considered in this paper. (a) and (b) are the $\v k$-vectors of SkX2
and SkX3 in two dimensions. (c), (d) and (e) are the $\v k$-vectors
of SC, FCC and BCC spin crystal states,
respectively.}\label{fig:k-vector}
\end{figure}

At $N=4$ one obtains face-centered-cubic (FCC) multiple spiral spin
structure\cite{binz} consisting of $\v k_1 = (k/\sqrt{3})(1, 1, 1)$,
$\v k_2 = (k/\sqrt{3})(-1, -1, 1)$, $\v k_3 = (k/\sqrt{3})(-1, 1,
-1)$, $\v k_4 = (k/\sqrt{3})(1, -1, -1)$, and the spin configuration

\ba
&& \v n^\mathrm{FCC}_{\v r} = \v n_0 + \sum_\alpha \Bigl( \v
n_\alpha^\mathrm{FCC} e^{i \v k_\alpha \cdot \v r} + ( \v
n_\alpha^\mathrm{FCC})^* e^{-i \v k_\alpha \cdot \v r} \Bigr), \nn
&& \v n_1^\mathrm{FCC} = {n_H \over 2}e^{i\theta_1} \left({\hat{x}
+\hat{y}- 2\hat{z} \over \sqrt{6}} + i{ {\hat{x} - \hat{y}} \over
\sqrt{2}}\right),\nn
&& \v n_2^\mathrm{FCC} = {n_H \over 2}e^{i\theta_2} \left({\hat{x}
+\hat{y}+ 2\hat{z} \over \sqrt{6}} + i{ {\hat{x} - \hat{y}} \over
\sqrt{2}}\right),\nn
&& \v n_3^\mathrm{FCC} = {n_H \over 2}e^{i\theta_3} \left({\hat{x}
+2\hat{y}+ \hat{z} \over \sqrt{6}} +i{ {\hat{z} - \hat{x}} \over
\sqrt{2}}\right),\nn
&& \v n_4^\mathrm{FCC} = {n_H \over 2}e^{i\theta_4} \left({2\hat{x}
+ \hat{y}+ \hat{z} \over \sqrt{6}} +i{ {\hat{y} - \hat{z}} \over
\sqrt{2}}\right). \ea

Finally, at $N=6$ one obtains body-centered-cubic (BCC) multiple
spiral states\cite{binz} consisting of $\v k_1 = (k/\sqrt{2})(1, 1,
0)$, $\v k_2 = (k/\sqrt{2})(1, -1, 0)$, $\v k_3 = (k/\sqrt{2})(1, 0,
1)$, $\v k_4 = (k/\sqrt{2})(-1, 0, 1)$, $\v k_5 = (k/\sqrt{2})(0, 1,
1)$, $\v k_6 = (k/\sqrt{2})(0, 1, -1)$, and the spin configuration

\ba && \v n^\mathrm{BCC}_{\v r} = \v n_0 + \sum_\alpha \Bigl( \v
n_\alpha^\mathrm{BCC} e^{i \v k_\alpha \cdot \v r} + ( \v
n_\alpha^\mathrm{BCC})^*  e^{-i \v k_\alpha \cdot \v r} \Bigr) , \nn
&& \v n_1^\mathrm{BCC} = {n_H \over 2}e^{i\theta_1} \left(\!\hat{z}
\!+\! i{ {\hat{y} \!-\! \hat{x}} \over \sqrt{2}}\!\right)\!, \nn
&& \v n_2^\mathrm{BCC} = {n_H \over 2}e^{i\theta_2} \left(\!\hat{z}
\!+\! i{ {\hat{y} \!+\! \hat{x}} \over \sqrt{2}}\!\right)\!,\nn
&& \v n_3^\mathrm{BCC} = {n_H \over 2}e^{i\theta_3} \left(\!\hat{y}
\!+\! i{ {\hat{x} \!-\! \hat{z}} \over \sqrt{2}}\!\right)\!, \nn
&& \v n_4^\mathrm{BCC} = {n_H \over 2}e^{i\theta_4} \left(\!\hat{y}
\!+\! i{ {\hat{x} \!+\! \hat{z}} \over \sqrt{2}}\!\right)\!, \nn
&& \v n_5^\mathrm{BCC} = {n_H \over 2}e^{i\theta_5} \left(\!\hat{x}
\!+\! i{ {\hat{z} \!-\! \hat{y}} \over \sqrt{2}}\!\right)\!, \nn
&& \v n_6^\mathrm{BCC} = {n_H \over 2}e^{i\theta_6} \left(\!\hat{x}
\!+\! i{ {\hat{z} \!+\! \hat{y}} \over \sqrt{2}}\!\right)\!. \ea

The last three multiple-spin configurations (SC, FCC, BCC) were
first considered by Binz \textit{et al.}\cite{binz} under the
general effort to understand the ``partial-order" phase of bulk MnSi
at high pressure\cite{partial-order}. Their method of analyzing the
interaction effects is different from ours, and the magnetic field
effect was not extensively discussed in their work.

By feeding all eight spin configurations into the free energy form
(\ref{eq:free-energy}) it becomes possible to evaluate their
energies as a function of $B$. The quadratic theory consisting of
$J$, $D$, and $\v B$ alone were already shown to give degenerate
energy, Eq. (\ref{eq:quadratic-E}). Evaluation of the various
anisotropy energies can be found in the Appendix and here we only
quote the final expression. The final energies are conveniently
expressed in the reduced units,

\ba {C \over J} \rightarrow C, ~ {U \over D^2/J }\rightarrow U, ~ {W
\over D^2/J } \rightarrow W, ~ { B \over D^2/J } \rightarrow B. \ea

\begin{widetext}

\ba \mathrm{H[111]}: && {1\over 2} {n_0^2-1\over 1+C/3} +U\Bigl(-{4
\over 3}n_0^4 + {4 \over 3}n_0^2 + 1\Bigr) + W\Bigl(-{1 \over
2}n_0^4 + n_0^2 + {1 \over 2}\Big)- B n_0, \nonumber \ea
\ba \mathrm{H[110]}:  && {1\over 2}{n_0^2-1\over 1+C/4}
+U\Bigl(-2n_0^4 + 2n_0^2 + 1\Bigr) + W\Bigl(-{23 \over 16}n_0^4 +
{15 \over 8}n_0^2 + {9 \over 16}\Big)- B n_0, \nonumber \ea
\ba \mathrm{Conical}: && {1\over 2}(n_0^2-1)  +U + W \Bigl({7\over
4}n_0^4 - {3\over 2}n_0^2 +{3\over 4}\Bigr) - B n_0, \nonumber \ea
\ba \mathrm{SkX2}: && {1 \over 2} (n_0^2 - 1) +U\Bigl(-{7 \over
4}n_0^4 + {3 \over 2} n_0^2 + {5 \over 4} \Bigl) + W \Bigl(-{5 \over
4}n_0^4 + {3 \over 2} n_0^2+ {3\over 4}\Bigr) -  B n_0, \nonumber
\ea
\ba \mathrm{SkX3}:  && {1\over 2}{n_0^2-1\over 1+C/8} + U \left(-9
n_0 \Bigl( {1-n_0^2\over 3}\Bigr)^{3/2}-{19 \over 12}n_0^4 +{7 \over
6} n_0^2 + {17 \over 12}\right) \nn
&&~~~~+ W \left(-6 n_0 \Bigl( {1-n_0^2\over 3}\Bigr)^{3/2} - {35
\over 32 }n_0^4 + {19 \over 16} n_0^2  +{29 \over 32}\right)- B n_0,
\nonumber \ea
\ba \mathrm{SC}: && {1 \over 2} (n_0^2 - 1) +U\Bigl(-n_0^4 + {2
\over 3} n_0^2 + {4 \over 3} \Bigl) + W \Bigl(-{1 \over 4}n_0^4 + {1
\over 2} n_0^2 + {3 \over 4} \Bigr) -  B n_0, \nonumber \ea
\ba \mathrm{FCC}:  && {1 \over 2}  {n_0^2 - 1 \over 1+C/3}
+U\Bigl(-{11 \over 12}n_0^4 - {1 \over 2} n_0^2 + {17 \over
12}\Bigr) + W\Bigl(-{1 \over 8}n_0^4 + {1 \over 4} n_0^2 +{7 \over
8}\Big)- B n_0, \nonumber \ea \ba \mathrm{BCC}: && {1 \over 2}
{n_0^2 - 1 \over 1+C/8} +U\Bigl({1 \over 6}n_0^4 -{2 \over 3} n_0^2
+ {3 \over 2} \Bigr)+ W\Bigl(-{3 \over 32}n_0^4 + {3 \over 32} n_0^2
+ {29 \over 32}\Bigr) \nn
&& ~~~ - 4\sqrt{27U^2 +30UW + 9W^2} n_0 \Bigl({1-n_0^2 \over
6}\Bigl)^{3/2} - B n_0. \label{eq:total-E}\ea
\end{widetext}
The normalization $n_0^2 + Nn_H^2=1$ is used to express the energies
solely in terms of $n_0$. The above energies can be minimized for
$n_0$ between $0$ and $1$ for a given field strength $B$, and such
energies can be compared among the eight candidate states. In
particular we are interested in the parameter regime $(C,U,W)$ in
which one of the multiple-spiral states becomes the ground state,
without invoking the finite-temperature effects as in previous
Ginzburg-Landau
calculations\cite{pfleiderer-1,bogdanov-nature,bogdanov-arxiv}.

\section{Ground State Phases}
\label{sec:phase-diagram}

\begin{figure}[ht]
\includegraphics[scale=0.3]{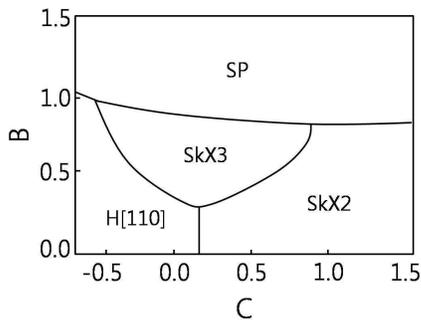}
\caption{Ground state phase diagram for two-dimensional spin states,
\{H[110], SkX2, SkX3\}. In a previous work\cite{YONH} only $C>0$ was
considered. Interaction parameters are $(U,W)=(0,0.1)$ for this
phase diagram. SP refers to full spin polarization, or ferromagnetic
state. }\label{fig:2D-phases}
\end{figure}

Previous two-dimensional lattice model study yielded a phase diagram
with the ground states H[110], SkX2, and SkX3 appearing over various
sectors of the $(C,B)$-plane\cite{YONH} (The parameter $A_2$ in Ref.
\onlinecite{YONH} corresponds to $C$ in the present model). Our
energy calculation in the continuum model reproduces qualitatively
the same phase diagram (Fig. \ref{fig:2D-phases}) when the search
for minimum energy was confined to three candidate states of
two-dimensional character, \{H[110], SkX2, SkX3\}.

\begin{figure}[ht]
\includegraphics[scale=0.4]{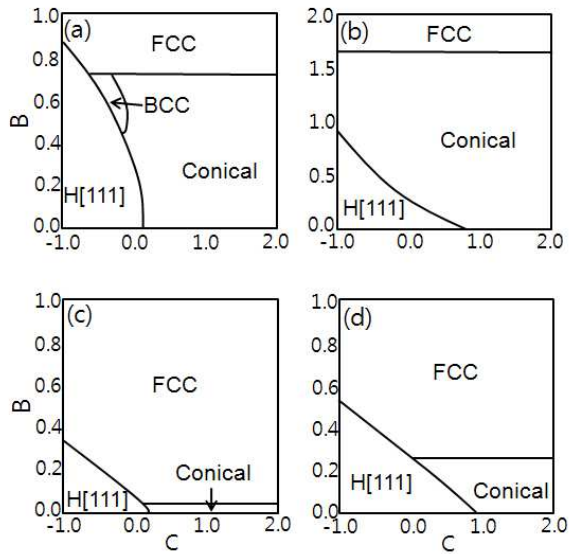}
\caption{Ground state phase diagram for all the states we
considered. (a) $(U,W)=(0.1,0.1)$, (b) $(U,W)=(0.1,0.5)$, (a)
$(U,W)=(0.5,0.1)$, (a) $(U,W)=(0.5,0.5)$.}\label{fig:3D-phases}
\end{figure}

When three single spiral states are compared in energy, either H[111]
or conical states win out, leaving no room for H[110] ground state
for any of the parameter ranges we searched. We conclude that H[110]
found in the thin-film chiral magnet\cite{tokura} was a special
feature of two dimensionality.

Now all the states considered in this paper are compared together
and the resulting phase diagram is shown in Fig.
\ref{fig:3D-phases}. SkX2 and SkX3 states have disappeared, replaced
by spiral states H[111] or conical at low field. Besides the $(U,W)$
values shown, we have searched for a wider set of parameters but
still did not find a region with SkX2, SkX3, or SC ground state.
Comparing Fig. \ref{fig:2D-phases} and Fig. \ref{fig:3D-phases}, we
conclude that the two-dimensional Skyrme crystal phases are stable
only when the corresponding spatial dimensionality is also two. As
shown in Fig. \ref{fig:3D-phases}, FCC state enjoys a particular
energetic advantage in the high-field region while the low-field
side is dominated by spiral states. A larger $U$ value tends to push
the FCC region down to a lower magnetic field [Fig.
\ref{fig:3D-phases} (c) and (d)].

\section{Skyrmions and Hedgehogs}
\label{sec:chirality}

In the previous section we have established the phase diagram of the
soft-spin model in the variational space of single and multiple
spiral phases. The main interest of this paper, we emphasize, is in
the identification of three-dimensional phases carrying a non-trivial
Skyrmion number. A general consideration relates the Skyrmion number
to the effective magnetic field felt by the underlying conduction
electrons and the emergence of anomalous Hall effect (sometimes
called topological Hall effect)\cite{spin-chirality}. With a
three-dimensional space one can define such number for each $xy$
plane along the $z$-axis, the direction of the applied field, as

\ba Q (z) = {1 \over 4 \pi }\int d^2 \v r ~ \v n \cdot
\Bigl({\partial \v n \over
\partial x}\times {\partial \v n \over \partial y}\Bigr). \label{eq:q}\ea
The integration extends over one unit cell in real space in the $xy$
plane. This quantity is an integer\cite{raj} for the unit-modulus
field satisfying $(\v n_{\v r} )^2 = 1$. Our soft-spin
configurations generally do not obey this condition locally, but
only on average, $\langle \v n_{\v r}^2 \rangle = 1$. Let's
therefore improve upon the above definition by using the normalized
vector field $\hat{\v n} = \v n /|\v n|$ to define the Skyrmion
number with proper topological characteristics as

\ba \hat{Q}(z) &=& {1 \over 4 \pi }\int d^2 \v r ~ \hat{\v n} \cdot
\Bigl({\partial \hat{\v n} \over
\partial x}\times {\partial \hat{\v n} \over \partial y}\Bigr) \nn
&& = {1 \over 4 \pi }\int d^2 \v r ~ {1\over \v n^3}\v n \cdot
\Bigl({\partial \v n \over
\partial x}\times {\partial \v n \over \partial y}\Bigr) .
\label{eq:q2}\ea
The second line follows from the first by a simple manipulation. In
the following, we will be largely concerned with the evaluation of
$\hat{Q}(z)$ in each of the multiple-spiral phases and how it
depends on the applied field through the uniform magnetization $\v
n_0$.

\subsection{SkX2 \& SkX3}
For the two-dimensional spin crystal SkX2 the spin configuration, up
to an overall constant, is given by

\ba \v n \propto ( \sin y , \cos x , \cos y + \sin x +m) .\ea
We choose $k \equiv 1$ and $m \equiv n_0 / n_H$ for convenience.
Actually one finds a periodic array of ``nodes" (localized points
where the magnetization vector $\v n$ vanishes) for $m = 0$.
Expansion around a particular such node $(x_0, y_0)$ gives the
linearized spin structure

\ba \v n \propto (y \cos y_0, -x \sin x_0, m -y \cos y_0 + x \sin
x_0 ).\ea
The coordinates are measured from the nodal position. It leads to
the Skyrmion density
\ba {1 \over 4 \pi} {P(m) \over [X^2 + Y^2 + P(m)^2 ]^{3/2}},
\label{eq:SkX2-Skyrme-density}\ea
where
\ba P(m) = {m \sin x_0 \cos y_0 \over \sqrt{1 - \cos^2 x_0 \sin^2
y_0} },\ea
and $(X, Y)$ is a linear coordinate transformation of $(x, y)$. The
node positions $(x_0, y_0)$ are $(-\pi/2, 0)$ and $(\pi/2,\pi)$ in a
unit cell of size $(2\pi)\times(2\pi)$, and both points give $P(m)
=-m$. Integrating Eq. (\ref{eq:SkX2-Skyrme-density}) over $(X, Y)$
gives the Skyrmion number

\ba \hat{Q}(m) = -{1 \over 2}\mathrm{sgn}[m]. \ea
As $m$ passes from positive to negative, the Skyrmion number changes
from $-1/2$ to $+1/2$. With two such nodes per unit cell, the total
Skyrmion number per unit cell depends on $m$ as $-\mathrm{sgn}(m)$.
It is convenient to view the situation in the enlarged
three-dimensional space consisting of two spatial coordinates
$(x,y)$ and one parameter $m$. The set of coordinates $(x, y, m) =
(x_0, y_0, 0)$, where $(x_0, y_0)$ is the node position of SkX2 at
$m = 0$, define the centers of ``hedgehogs". As is well known, a
passage through the hedgehog center changes the Skyrmion number by
one unit.

One can carry out a similar analysis for SkX3, for which the spin
structure reads

\ba \v n &\propto& (n_x, n_y, n_z), \nn
 n_x &=& \sqrt{3} \cos {x \over 2}
\sin {\sqrt{3} y \over 2} , \nn
n_y &=& -\Bigl( 2 \cos {x \over 2} + \cos {\sqrt{3} y \over 2}
\Bigr) \sin {x \over 2},\nn
n_z &=& \cos x + 2 \cos {x \over 2} \cos {\sqrt{3} y \over 2} + m
 .\ea
There are no nodes in this magnetization profile, and by direct
evaluation one finds $\hat{Q}(z) = -1$ for all uniform magnetization
$-1 < m < 1$. In a sense, SkX3 is a unique phase carrying a
nontrivial Skyrmion number without possessing (or even proximate to)
real-space singularities in the spin structure. Indeed, recent
experiments\cite{AHE-2,tokura} single out the SkX3 phase as the only
topologically non-trivial magnetic phase so far discovered in chiral
magnets.

One easily realizes that a node in the magnetization is a rare event
in the two-dimensional spin structure. Each component of the spin
vector taken equal to zero define a two-dimensional surface embedded
in a three-dimensional space. For 2D spin structures such as SkX2
and SkX3, such a surface is extended vertically along the $z$-axis
and traces out a curve in the $xy$ plane. The interaction of three
curves, from each component of the spin vector taken to zero,
defines the node. The three curves in a plane do not cross at a
single point except through fine tuning, such as the tuning of $n_0$
through zero in the SkX2 case. With both SkX2 and SkX3, the nodeless
spin structures also happen to be topologically nontrivial and carry
a finite Skyrmion number.

The same reasoning applied to a generic three-dimensional spin
crystal structure states that nodes are unavoidable because three
independent surfaces would cross at isolated points. Below we show
that each such node is the center of a hedgehog or an anti-hedgehog,
a passage through which the Skyrmion number changes by $+1 (-1)$.

\begin{center}
\begin{table}[t]
\caption{Nodes of the magnetization vector in a unit cell for SC.
$P(z,m)$, defined in Eq. (\ref{eq:Pzm-for-SC}), associated with each
zero can tell whether a given node is the center of a hedgehog or an
an anti-hedgehog ($\eta = \sin^{-1}(m/\sqrt{2}$)).}
\begin{tabular}{c|c|c|c}
\hline \hline
$x_0$ & $y_0$ & $z_0$ & $P(z,m)$\\ [0.5ex]
\hline
$-{3\pi \over 4} + \eta$ & ${\pi \over 4} + \eta$ & ${3\pi \over 4}
+ \eta $ & $\sqrt{2-m^2}(1-m^2)z/2$
\\[0.5ex]
${3\pi \over 4} + \eta$ & ${3\pi \over 4} + \eta$ & ${3\pi \over 4}
- \eta $ & $\sqrt{2-m^2}(m^2 -1)z/2$
\\[0.5ex]
${3\pi \over 4} + \eta$ & $-{3\pi \over 4} - \eta$ & ${\pi \over 4}
+ \eta $ & $\sqrt{2-m^2}(1-m^2)z/2$
\\[0.5ex]
$-{3\pi \over 4} + \eta$ & $-{\pi \over 4} - \eta$ & ${\pi \over 4}
- \eta $ & $\sqrt{2-m^2}(m^2 -1)z/2$
\\[0.5ex]
$-{\pi \over 4} - \eta$ & $-{\pi \over 4} - \eta$ & $-{\pi \over 4}
+ \eta $ & $\sqrt{2-m^2}(1-m^2)z/2$
\\[0.5ex]
${\pi \over 4} - \eta$ & $-{3\pi \over 4} - \eta$ & $- {\pi \over 4}
- \eta $ & $\sqrt{2-m^2}(m^2 -1)z/2$
\\[0.5ex]
${\pi \over 4} - \eta$ & ${3\pi \over 4} + \eta$ & $-{3\pi \over 4}
+ \eta $ & $\sqrt{2-m^2}(1-m^2)z/2$
\\[0.5ex]
$-{\pi \over 4} - \eta$ & ${\pi \over 4} + \eta$ & $-{3\pi \over 4}
- \eta $ & $\sqrt{2-m^2}(m^2 -1)z/2$
\\[0.5ex]
\hline \hline
\end{tabular}
\label{table:zeros-of-SC}
\end{table}
\end{center}

\subsection{SC, FCC \& BCC}
\begin{figure}[h]
\includegraphics[scale=0.4]{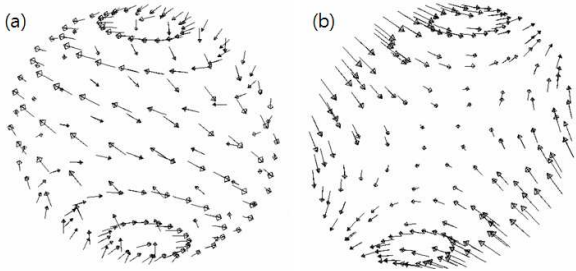}
\caption{Spin configuration around (a) hedgehog and (b) anti-hedgehog
center in the SC spin crystal. Magnetization vector $\v n$ vanishes
at the center of each singularity.}\label{fig:SC-chirality}
\end{figure}

Taking $k=1$, the spin configuration for SC becomes

\ba \v n \propto (\sin y \!+\! \cos z, \sin z \!+\! \cos x, \sin x
\!+\! \cos y \!+\! m ) .\ea
The three phase factors $\theta_\alpha$ are all assumed to be zero.
There are eight independent zeroes in a unit cell and their
coordinates are listed in Table \ref{table:zeros-of-SC}.
Taylor-expanding around a given node $(x_0, y_0, z_0)$ to first
order in the deviation results in

\ba \v n \!\propto\! (y \cos y_0  \!-\! z \sin z_0  ,  z \cos
z_0\!-\!x \sin x_0 ,  x\cos x_0 \!-\! y \sin y_0 ) . \nn
\label{eq:hedgehog}\ea
The coordinates $(x,y,z)$ are measured from their zeroes.  The triple
product $\v n \cdot \partial_x \v n \times \partial_y \v n = z (\cos
x_0 \cos y_0 \cos z_0 - \sin x_0 \sin y_0 \sin z_0 )$ follows
immediately, and the normalized Skyrmion density becomes

\ba {1 \over 4\pi} {P(z,m) \over [ X^2 + Y^2 + P(z,m)^2 ]^{3/2}}
\nonumber \ea
where

\ba P(z,m) = {z (\cos x_0 \cos y_0 \cos z_0 - \sin x_0 \sin y_0 \sin
z_0) \over \sqrt{1 - \cos^2 x_0 \sin^2
y_0}},\nn\label{eq:Pzm-for-SC}\ea
and $(X,Y)$ is again related to $(x,y)$ through linear coordinate
change. Integral over $(X,Y)$ yields the Skyrmion number

\ba \hat{Q}(z) = {1\over 2} \mathrm{sgn}[ P(z,m) ], \ea
with $P(z,m)$ for each node listed in Table \ref{table:zeros-of-SC}.
We define such node with $\hat{Q}(z=0^+)-\hat{Q}(z=0^-) = +1$ as the
hedgehog center, and the other one with
$\hat{Q}(z=0^+)-\hat{Q}(z=0^-) = -1$ as the anti-hedgehog center. A
snapshot of a pair of hedgehog/anti-hedgehog configurations obtained
from Eq. (\ref{eq:hedgehog}) is shown in Fig.
\ref{fig:SC-chirality}.

As can be seen from Table \ref{table:zeros-of-SC}, hedgehogs and
anti-hedgehogs occur in alternating fashion along the $z$-axis.
Average over a unit cell in the $z$-direction yields the average
Skyrmion number $\overline{Q} = \int_0^{2\pi} dz \hat{Q}(z) / 2\pi$
which equals $(\eta\equiv \sin^{-1}(m/\sqrt{2})$

\ba \overline{Q} &=& \left\{
\begin{array}{ll}
 -{4 \over \pi } \eta & \left(0<\eta<{\pi \over 4} \right) \\
  {4 \over \pi }\left(\eta-{\pi \over 2} \right) & \left({\pi\over 4}<\eta<{\pi
\over 2}\right)
\end{array} \right.  \nn
\eta &\equiv& \sin^{-1}\bigl(m/\sqrt{2}\bigr)
.\label{eq:Qaverage-SC} \ea
When $m$ exceeds $\sqrt{2}$, $\hat{Q}(z)$ is
identically zero.

The FCC spin configuration can be analyzed with the similar strategy
after taking $k=\sqrt{3}$. Keeping track of the node positions
proved difficult for general phase angles, but we do achieve some
simplification with special choices, \textit{e.g.} $\theta_1 =
\theta_2 = 0$ and $\theta_3 = - \theta_4 = -\pi / 3$. In this case
nodes occurs at

\ba \cos z \Bigl( \cos (x \!+\! y) \!- \!\sqrt{3} \sin (x \!-\! y)
\Bigr) &=& 0,\nn
\sin z \Bigl( \sqrt{3} \cos ( x \!+\! y ) \!-\! \sin (x
\!-\!y)\Bigr) &=& 0, \nn
2 \sin z \sin (x\!+\! y) \!+\! 2\cos z \cos (x \!-\! y)  \!+\! m &=&
0.\ea
One possible set of zeroes come from choosing $\cos z = 0$, $\sqrt{3}
\cos ( x \!+\! y ) \!=\! \sin (x \!-\!y)$ and $2 \sin z\sin (x\!+\!
y) \!+\! m =0$. There are eight possible solutions. Another possible
set is $\sin z = 0$, $\cos (x \!+\! y) \!= \!\sqrt{3} \sin (x \!-\!
y)$ and $2 \cos z\cos (x \!-\! y) \!+\! m =0$. This gives another
eight zeroes. A third set arises from having $\cos (x \!+\! y)=0$,
$\sin(x \!-\! y)=0$ and $\sin z \Bigl( \sqrt{3} \cos ( x \!+\! y )
\!-\! \sin (x \!-\!y)\Bigr) = 0$. There are 16 independent nodes that
belong to the third set as listed in Table \ref{table:zeros-of-FCC}.
From the Table one can read off the $z$-integrated Skyrmion number.
The result is in fact twice that given Eq. (\ref{eq:Qaverage-SC}) for
SC if one replaces $\eta=\sin^{-1}(m/\sqrt{2})$ with $\xi =
\sin^{-1}(m/2\sqrt{2})$ now. The first two sets of nodes contribute
zero Skyrmion number as each $z$-plane contains an equal number of
hedgehogs and anti-hedgehogs. For BCC spin crystal phase Binz and
Vishwanath obtained $\overline{Q} = - \mathrm{sgn}(m)(2/\pi)
\cos^{-1} (|m|/2\sqrt{2})$ for $|m|<2\sqrt{2}$ and zero
otherwise\cite{binz-physica}. Together with our calculation, a
complete information is available for the magnetization (magnetic
field dependence) of the Skyrmion number for bulk spin crystal
phases.

\begin{center}
\begin{table}
\caption{Nodes of the magnetization vector in a unit cell for FCC
($\xi = \sin^{-1}(m / 2\sqrt{2})$)).}
\begin{tabular}{c|c|c|c}
\hline \hline
$x_0$ & $y_0$ & $z_0$ & $P(z,m)$\\ [0.5ex]
\hline
${\pi / 4} $ & ${5\pi / 4} $ & ${7\pi / 4} + \xi $ &
$\sqrt{8-m^2}(4-m^2)z$
\\[0.5ex]
${5\pi / 4} $ & ${\pi / 4} $ & ${7\pi / 4} + \xi $ &
$\sqrt{8-m^2}(4-m^2)z$
\\[0.5ex]
${\pi / 4} $ & ${\pi / 4} $ & ${7\pi / 4} - \xi $ &
$\sqrt{8-m^2}(m^2-4)z$
\\[0.5ex]
${5\pi / 4} $ & ${5\pi / 4} $ & ${7\pi / 4} - \xi $ &
$\sqrt{8-m^2}(m^2-4)z$
\\[0.5ex]
${3\pi / 4} $ & ${7\pi / 4} $ & ${5\pi / 4} + \xi $ &
$\sqrt{8-m^2}(4-m^2)z$
\\[0.5ex]
${7\pi / 4} $ & ${3\pi / 4} $ & ${5\pi / 4} + \xi $ &
$\sqrt{8-m^2}(4-m^2)z$
\\[0.5ex]
${3\pi / 4} $ & ${3\pi / 4} $ & ${5\pi / 4} - \xi $ &
$\sqrt{8-m^2}(m^2-4)z$
\\[0.5ex]
${7\pi / 4} $ & ${7\pi / 4} $ & ${5\pi / 4} - \xi $ &
$\sqrt{8-m^2}(m^2-4)z$
\\[0.5ex]
${\pi / 4} $ & ${\pi / 4} $ & ${3\pi / 4} + \xi $ &
$\sqrt{8-m^2}(4-m^2)z$
\\[0.5ex]
${5\pi / 4} $ & ${5\pi / 4} $ & ${3\pi / 4} + \xi $ &
$\sqrt{8-m^2}(4-m^2)z$
\\[0.5ex]
${\pi / 4} $ & ${5\pi / 4} $ & ${3\pi / 4} - \xi $ &
$\sqrt{8-m^2}(m^2-4)z$
\\[0.5ex]
${5\pi / 4} $ & ${\pi / 4} $ & ${3\pi / 4} - \xi $ &
$\sqrt{8-m^2}(m^2-4)z$
\\[0.5ex]
${3\pi / 4} $ & ${3\pi / 4} $ & ${\pi / 4} + \xi $ &
$\sqrt{8-m^2}(4-m^2)z$
\\[0.5ex]
${7\pi / 4} $ & ${7\pi / 4} $ & ${\pi / 4} + \xi $ &
$\sqrt{8-m^2}(4-m^2)z$
\\[0.5ex]
${3\pi / 4} $ & ${7\pi / 4} $ & ${\pi / 4} - \xi $ &
$\sqrt{8-m^2}(m^2-4)z$
\\[0.5ex]
${7\pi / 4} $ & ${3\pi / 4} $ & ${\pi / 4} - \xi $ &
$\sqrt{8-m^2}(m^2-4)z$
\\[0.5ex]
\hline \hline
\end{tabular}
\label{table:zeros-of-FCC}
\end{table}
\end{center}

\begin{figure}[ht]
\includegraphics[scale=0.4]{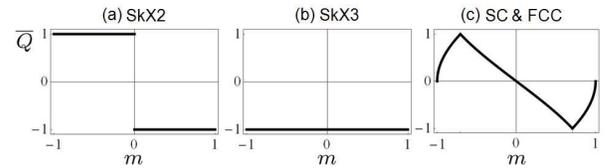}
\caption{$z$-integrated Skyrmion number $\overline{Q}$ for (a) SkX2,
(b) SkX3, and (c) SC and FCC spin crystals. The FCC result is
multiplied by $1/2$ in the figure. Magnetization $m$ is normalized
to unity at the maximum value with nonzero $\overline{Q}$.
}\label{fig:Q-vs-m}
\end{figure}

\subsection{Nature of 3D Spin Crystals}
Based on the analysis of the multiple-spiral states both of 2D and 3D
characters, we are led to suspect that realizing a topologically
non-trivial (in the sense of having nonzero $\hat{Q}(z)$) spin
structure free of singularities is extremely difficult in three
dimensions, whereas it should be quite easy in two dimensions. A
genuine 3D Skyrmion of course is without a singularity\cite{Skyrme},
and only a lattice of such structures would qualify as a  ``3D Skymre
crystal"\cite{klebanov}. All the 3D states discussed in this paper
and elsewhere\cite{binz} can be instead described as a ``hedgehog
crystal".

In the hard-spin limit where $\v n^2 =1$ is rigorously enforced at
the local level, we will find that a periodic array of singularities
becomes energetically forbidding and give way to single-spiral
structures without nodes. It is possible, however, that the spins
remain soft in real chiral magnets due to interaction with the
metallic host that can ``absorb" some of the spin from the local
moments\cite{bogdanov-nature}. In such a case, the singularities
inherent in the hedgehog lattice can be realized without an excessive
cost in energy. The other exciting possibility is that the spins
remain hard even against the metallic background, and yet somehow
form a topologically non-trivial 3D Skyrme crystal phase such as
envisioned by Klebanov and others\cite{klebanov}. Whether such a
state can be energetically stable compared to hedgehog or
single-spiral phases is a subject yet to be explored in the context
of chiral magnetism. Creating a singularity-free Skyrmion lattice
structure requires the inclusion of higher-order harmonics (in the
sense of Fourier decomposition of the spin structure), which also
partly explains why our first-harmonic construction only yields
hedgehog lattices.

\section{Summary}
\label{sec:summary}

The ground state energies of several single- and multiple-spiral
spin configurations are considered, on the basis of the
Ginzburg-Landau free energy suggested by Bak and Jensen\cite{bak}.
Phase diagrams in the space of anisotropy and magnetic field $(C,B)$
are worked out for a variety of interaction parameters $(U,W)$.
Previous two-dimensional phase diagram\cite{YONH} is recovered and
three-dimensional phase diagrams are worked out anew. Similar
efforts were made earlier by Binz \textit{et al.}\cite{binz} to
carve out the 3D phase diagram, focusing on the stability of 3D spin
textures such as SC, FCC, and BCC, under zero magnetic field and
using the interaction terms which are non-local in space. Leonov
\textit{et al.}\cite{bogdanov-arxiv} considered the phase diagram
spanned by temperature and field, with two-dimensional spin textures
SkX2 and SkX3 as candidate states. Our search is confined to
zero-temperature, but covers both 2D and 3D spin textures.

As shown in Fig. \ref{fig:3D-phases}, magnetic field plays a crucial
role in stabilizing Skyrmionic order over single-spiral states in
both 2D and 3D geometries. Generically, two-dimensional Skyrme
crystals (SkX2 and SkX3) are favored only if the corresponding
effective space dimensionality is two. For three dimensions, 3D
Skyrmion structures such as FCC and BCC are more favorable
energetically. Our calculation suggests the possibility for a spin
crystal phase of FCC symmetry very close to zero magnetic field for
a sufficiently strong interaction parameter $U$.

The Skyrmion number embodied by each multiple spiral phase was
examined. An array of hedgehog and anti-hedgehog centers were
identified in the case of three-dimensional spin crystals. Passage
through each such center change the Skyrmion number by $\pm 1$.
Generally there exists a non-zero, $z$-integrated Skyrmion number
for finite uniform magnetization. We showed that SkX3 is the only
spin crystal structure without singularities and yet exhibiting a
nonzero Skyrmion number. Constructing a singularity-free spin
crystal state in three-dimensional chiral magnet is an interesting
challenge for the future.

\acknowledgments  H. J. H. is supported by Mid-career Researcher
Program through NRF grant funded by the MEST (No.
R01-2008-000-20586-0), and in part by the Asia Pacific Center for
Theoretical Physics. H. J. H. is grateful to Naoto Nagaosa and Jan
Zaanen for their invaluable inputs on Skyrmions and hedgehogs.

\appendix
\section{Calculation of $C$-anisotropy}
To calculate the $C$-anisotropy one should first obtain $\partial
n_x /\partial x$:

\ba {\partial n_x\over \partial x} &=& \sum_{\alpha} \Bigl(
ik_\alpha^x n_{\alpha}^x e^{i \v k_{\alpha} \cdot \v r} - ik_\alpha^x
(n_{\alpha}^x)^* e^{-i \v k_{\alpha} \cdot \v r}\Bigr),\nn
\int \Bigl({\partial n_x\over \partial x}\Bigr)^2 &=& 2 \sum_{\alpha}
(k_\alpha^x)^2 n_{\alpha}^x (n_{\alpha}^x)^*. \ea
With similar expressions from $\int (\partial n_y/\partial y)^2$ and
$\int (\partial n_z /\partial z)^2$, $C$-anisotropy energy reads
\ba &&{1 \over 2} C \int \left[\Bigl({\partial n_x \over
\partial x}\Bigr)^2 + \Bigl({\partial n_y \over \partial y}\Bigr)^2
+ \Bigl({\partial n_z \over \partial z}\Bigr)^2 \right]  \!=\! \nn
&& C \sum_{\alpha}\!\Bigl(\!(k_\alpha^x)^2 n_{\alpha}^x
(n_{\alpha}^x)^* \!\!+\!\!(k_\alpha^y)^2 n_{\alpha}^y
(n_{\alpha}^y)^* \!\!+\!\! (k_\alpha^z)^2 n_{\alpha}^z
(n_{\alpha}^z)^*\!\Bigr)\!.\nn \ea
The eight states have the $C$-anisotropy energies

\ba
\mathrm{Helical}(\mathrm{H}_{111}): & & (1/6)C n_H^2 k^2 \nn
\mathrm{Helical}(\mathrm{H}_{110}): & & (1/8) C n_H^2 k^2 \nn
\mathrm{Conical}: & & 0\nn
\mathrm{SkX2}: && 0 \nn
\mathrm{SkX3}: && (3/16) C n_H^2 k^2 \nn
\mathrm{SC}: && 0 \nn
\mathrm{FCC}: && (2/3) C n_H^2 k^2 \nn
\mathrm{BCC}: && (3/8) C n_H^2 k^2 \ea
Including the $C$-anisotropy gives the total energy for the various
spin states

\ba \mathrm{H_{111}}:  && {1 \over 2}n_H^2 \Bigl(J  k^2 - 2 D k\Bigr)
+{ 1\over 6}C n_H^2 k^2 \nn
\mathrm{H_{110}}: && {1 \over 2}n_H^2 \Bigl(J  k^2 -2 D  k \Bigr)+{
1\over 8}C n_H^2 k^2 \nn
\mathrm{Co}: && {1 \over 2}n_H^2 \Bigl(J  k^2 - 2 D k\Bigr)\nn
\mathrm{SkX2}: && n_H^2 \Bigl(J  k^2 -2 D  k \Bigr) \nn
\mathrm{SkX3}: && {3 \over 2} n_H^2 \Bigl(J  k^2 -2 D  k
\Bigr)+{3\over 16}C n_H^2 k^2 \nn
\mathrm{SC}: && {3 \over 2}n_H^2 \Bigl(J  k^2 -2 D  k \Bigr) \nn
\mathrm{FCC}:  && 2 n_H^2 \Bigl(J  k^2 -2 D  k \Bigr)+{ 2\over 3}C
n_H^2 k^2\nn
\mathrm{BCC}: && 3 n_H^2 \Bigl(J  k^2 -2 D  k \Bigr)+{ 3\over 8}C
n_H^2 k^2 \ea
As the remaining $U$ and $W$ interactions do not contain the
$k$-dependence, one can minimize $k$ for the above energies and
obtain the optimal $\v k$-vector length for each spin state

\ba k^\mathrm{H_{111}}= D/(J\!+\! {C/3}), && k^\mathrm{H_{110}} =
D/(J \!+\! {C /4}),\nn
k^\mathrm{Co} = D/J, && k^\mathrm{SkX2} = D/J ,\nn
k^\mathrm{SkX3} = D/(J\!+\!{C/8}), && k^\mathrm{SC} =D/J,\nn
k^\mathrm{FCC} =D/(J\!+\!C/3), && k^\mathrm{BCC} =D/(J\!+\!C/8). \ea

\section{Calculation of $U$-energy}
Dividing up the spin configuration $\v n = \v n_0 + \v \Phi$ as in
Eq. (\ref{eq:n-alpha}),  $\v n^4$ is expanded as
\ba \v n^4 &=& (\v n_0^2 + 2 \v n_0 \cdot \v \Phi + \v \Phi^2 )^2
\nn
&=& \v n_0^4 + 4 (\v n_0 \cdot \v \Phi)^2 + 4 \v n_0^2(\v n_0 \cdot
\v \Phi) \nn
&& + 4 \v \Phi^2 (\v n_0 \cdot \v \Phi) + 2 \v n_0^2 \v \Phi^2 + \v
\Phi^4.\ea
H[111] state gives $\int (\v n_0 \cdot \v \Phi)^2 = n_0^2 n_H^2 /3
$, $\int \v \Phi^2 = n_H^2 $, $\int \v \Phi^4 = n_H^4$, and the
$U$-energy
\ba \mathrm{H[111]}: ~ U\left(n_0^4 +{10 \over 3} n_0^2n_H^2 +
n_H^4\right).\ea
H[110] state gives $\int (\v n_0 \cdot \v \Phi)^2 = n_0^2 n_H^2 /2$,
$\int \v \Phi^2 = n_H^2 $,  $\int \v \Phi^4 = n_H^4$, and the
$U$-energy
\ba \mathrm{H[110]}: ~  U\left(n_0^4 + 4 n_0^2n_H^2 +
n_H^4\right).\ea
Conical state gives $\int (\v n_0 \cdot \v \Phi)^2 = 0$, $\int \v
\Phi^2 = n_H^2 $, $\int \v \Phi^4 = n_H^4$, and the $U$-energy
\ba \mathrm{Conical}: ~ U\left(n_0^4 +2 n_0^2n_H^2 +
n_H^4\right).\ea

SkX2 state with $\int (\v n_0 \cdot \v \Phi)^2 = n_0^2 n_H^2$, $\int
\Phi^2 = 2 n_H^2$, and $\int \v \Phi^4 = 5 n_H^4$ gives the
$U$-energy

\ba \mathrm{SkX2}:~  U \left(n_0^4 + 8 n_0^2 n_H^2 +5 n_H^4\right).
\ea
There are four types of terms contributing to $\int \v \Phi^4$.

\begin{itemize}

\item  $\sum_\alpha  (\v n_\alpha \cdot \v n_\alpha^*)(\v
n_\alpha \cdot \v n_\alpha^*) = 2 n_H^4$

\item  $\sum_{l \neq m} (\v n_l \cdot \v n_l^*)(\v n_m \cdot \v
n_m^*) = 2 n_H^4$

\item $\sum_{l \neq m} (\v n_l \cdot \v n_m)(\v n_l^* \cdot \v
n_m^*) = (1/2) n_H^4$

\item $\sum_{l \neq m} (\v n_l \cdot \v n_m^*)(\v n_l^* \cdot \v
n_m) = (1/2) n_H^4$
\end{itemize}

SkX3 state gives $\int (\v n_0 \cdot \v \Phi)^2 = (3/2) n_0^2 n_H^2$
and $\int \v \Phi^2(\v n_0 \cdot \v \Phi)$ is non-vanishing,

\ba \int \v \Phi^2(\v n_0 \cdot \v \Phi) &=& \sum_{\v k_1\!+\!\v
k_2\!+\!\v k_3 \!=\! \v 0} (\v n_0 \cdot \v n_{\v k_1})(\v n_{\v k_2}
\cdot \v n_{\v k_3})\nn
& = &9 n_0 n_H^3 \cos(\theta_1 \!+\! \theta_2 \!+\! \theta_3) .\ea
There are 24 combinations of the type

\ba \int (\v n_0 \cdot \v n_1)(\v n_2 \cdot \v n_3) ={3 \over 16}n_0
n_H^3 e^{i(\theta_1 + \theta_2 + \theta_3)} \ea
and another 24 from their complex conjugates contributing to $\int \v
\Phi^2(\v n_0 \cdot \v \Phi)$. $\int \v \Phi^4$ gets contributions
from the following four terms.

\begin{itemize}

\item $\sum_\alpha  (\v n_\alpha \cdot \v n_\alpha^*)(\v n_\alpha
\cdot \v n_\alpha^*) = 3 n_H^4$

\item $\sum_{l \neq m} (\v n_l \cdot \v n_l^*)(\v n_m \cdot \v
n_m^*) = 6 n_H^4$

\item $\sum_{l \neq m} (\v n_l \cdot \v n_m)(\v n_l^* \cdot \v
n_m^*) = (27/8) n_H^4$

\item $\sum_{l \neq m} (\v n_l \cdot \v n_m^*)(\v n_l^* \cdot \v
n_m) = (3/8) n_H^4$

\end{itemize}
Overall, SkX3 has the $U$-energy
\ba \mathrm{SkX3}: ~ U\Bigl(n_0^4 \!+\! 9 n_0 n_H^3 \cos(\theta_1
\!+\! \theta_2 \!+\! \theta_3) \nn
\!+\! 12 n_0^2 n_H^2 \!+ {51 \over 4}n_H^4\Bigr) .\ea
The phase angles influence the total energy as  $\sim \cos (\theta_1
\!+\! \theta_2 \!+\! \theta_3)$. It can be minimized by choosing
$\theta_1 \!+\! \theta_2 \!+\! \theta_3 = \pi$.

SC state gives $\int (\v n_0 \cdot \v \Phi)^2 = n_0^2 n_H^2$ and
$\int \v \Phi^2 = 3n_H^2 $.  $\int \v \Phi^4 = 12 n_H^4 $ is
obtained from

\begin{itemize}
\item $\sum_\alpha (\v n_\alpha \cdot \v n_\alpha^*)(\v n_\alpha
\cdot \v n_\alpha^*) = 3 n_H^4$

\item $\sum_{l \neq m}
(\v n_l \cdot \v n_l^*)(\v n_m \cdot \v n_m^*) = 6 n_H^4$

\item $\sum_{l \neq m}(\v n_l \cdot \v n_m)(\v n_l^* \cdot \v
n_m^*) = (3/2) n_H^4$

\item $\sum_{l \neq m} (\v n_l
\cdot \v n_m^*)(\v n_l^* \cdot \v n_m) = (3/2) n_H^4$

\end{itemize}
$U$-energy for SC reads

\ba \mathrm{SC}: ~ U\left(n_0^4 +10 n_0^2n_H^2 +12 n_H^4\right). \ea

FCC state gives $\int (\v n_0 \cdot \v \Phi)^2 = 4n_0^2 n_H^2 /3 $,
and $\int \v \Phi^2 = 4n_H^2 $. $\int \v \Phi^4 = (68/3)n_H^4 $
derives from

\begin{itemize}
\item $\sum_\alpha  (\v n_\alpha \cdot \v n_\alpha^*)(\v n_\alpha
\cdot \v n_\alpha^*) = 4 n_H^4$

\item $\sum_{l \neq m} (\v n_l \cdot \v n_l^*)(\v n_m \cdot \v
n_m^*) = 12 n_H^4$

\item $\sum_{l \neq m} (\v n_l \cdot \v n_m)(\v n_l^* \cdot \v
n_m^*) = (8/3) n_H^4$

\item $\sum_{l \neq m} (\v n_l \cdot \v n_m^*)(\v n_l^* \cdot \v
n_m) = 4 n_H^4$

\end{itemize}
A careful calculation shows that contributions from all four wave
vectors different give zero contribution to $\int \v \Phi^4$. The
$U$-energy reads

\ba \mathrm{FCC}: ~ U\left(n_0^4 +{40 \over 3} n_0^2n_H^2 +{68 \over
3} n_H^4\right).\ea

BCC state gives $\int (\v n_0 \cdot \v \Phi)^2 = 2 n_0^2 n_H^2$ and
$\int \v \Phi^2 = 6 n_H^2 $. $\int \v \Phi^4 = 54 n_H^4 $ consists
of the following terms.

\begin{itemize}

\item $\sum_\alpha  (\v n_\alpha \cdot \v n_\alpha^*)(\v n_\alpha
\cdot \v n_\alpha^*) = 6 n_H^4$

\item $\sum_{l \neq m}(\v n_l \cdot \v n_l^*)(\v n_m \cdot \v
n_m^*) = 30 n_H^4$

\item $\sum_{l \neq m} (\v n_l \cdot \v n_m)(\v n_l^* \cdot \v
n_m^*) = 9 n_H^4$

\item $\sum_{l \neq m} (\v n_l \cdot \v n_m^*)(\v n_l^* \cdot \v
n_m) = 9 n_H^4$

\end{itemize}
As in SkX3,

\ba \int \v \Phi^2(\v n_0 \cdot \v \Phi)= \sum_{\v q_1 \!+\! \v q_2
\!+\! \v q_3\! =\! \v 0} (\v n_0 \cdot \v n_{\v q_1})(\v n_{\v q_2}
\cdot \v n_{\v q_3})\ea
is non-vanishing from combination of vectors $\v k_1 \!+\! \v k_4
\!-\! \v k_5 \!=\! \v 0$, $\v k_3 \!+\! \v k_6 \!-\! \v k_1 \!=\! \v
0$, $\v k_2 \!+\! \v k_5\!-\! \v k_3 \!=\! \v 0$, $\v k_2 \!+\! \v
k_4 \!+\! \v k_6 \!=\! \v 0$. The $U$-energy becomes

\ba && \mathrm{BCC}: ~  U\Bigl(n_0^4 + 14 n_0^2 n_H^2 + 54
n_H^4\Bigr) \nn
&& ~~~~~~ + U n_0 n_H^3\Bigl(\nn
&& 5  \cos(\theta_1 \!+\! \theta_4\!-\! \theta_5) \!-\!\sqrt{2}
\sin(\theta_1\!+\! \theta_4\!-\! \theta_5) \nn
&&+ 5  \cos(\theta_3\!+\! \theta_6\!-\! \theta_1) \!-\!\sqrt{2}
\sin(\theta_3\!+\! \theta_6\!-\! \theta_1)\nn
&& - 5 \cos(\theta_2\!+\! \theta_5\!-\! \theta_3) \!+\!\sqrt{2}
\sin(\theta_2\!+\! \theta_5\!-\! \theta_3)\nn
&&- 5 \cos(\theta_2\!+\! \theta_4\!+\! \theta_6) \!-\!\sqrt{2}
\sin(\theta_2\!+\! \theta_4\!+\! \theta_6)\Bigr). \ea
The phase angle dependence of the $U$-energy is more complicated than
in the case of SkX3 because there are six such angles instead of
three.

\section{Calculation of $W$-energy}
Spin  components are decomposed as $n_x = \Phi_x$, $n_y = \Phi_y$ and
$n_z = n_0 + \Phi_z$. Hence, $n_x^4=\Phi_x^4, n_y^4 = \Phi_y^4$, and

\ba n_z^4 &=& (n_0^2 + 2 n_0^2 \Phi_z^2 + \Phi_z^2)^2 \nn
&=& n_0^4 + 6 n_0^2 \Phi_z^2 + 4 n_0^3 \Phi_z + 4 n_0 \Phi_z^3
+\Phi_z^4. \ea
For all the states under consideration we have $\int n_x^4 = \int
n_y^4$.

H[111] state gives $\int n_x^4 = (1/6)n_H^4$ and $\int n_z^4 = n_0^4
+2n_0^2n_H^2 + (1/6)n_H^4$ for the $W$-energy

\ba \mathrm{H[111]}: ~ W\! \Bigl(n_0^4 \!+\! 2n_0^2 n_H^2 \!+\! {1
\over 2}n_H^4 \Bigr).\ea
H[110] state gives $\int n_x^4 = (3/32) n_H^4$ and $\int n_z^4 =
n_0^4 +3n_0^2n_H^2 + (3/8)n_H^4$ for the $W$-energy

\ba \mathrm{H[110]}: ~  W\Bigl(n_0^4 \!+\! 3n_0^2 n_H^2 \!+\! {9
\over 16}n_H^4 \Bigr).\ea
Conical state gives $\int n_x^4 = (3/8)n_H^4$ and $\int n_z^4 =
n_0^4$ for the $W$-energy

\ba \mathrm{Conical}: ~ W\Bigl({3 \over 4}n_H^4 + n_0^4\Bigr).\ea
SkX2 state has two types of terms in calculating $\int \Phi_z^4=(9/4)
n_H^4$.

\begin{itemize}

\item $\sum_\alpha  | (n_\alpha)_z |^4 = (3/4) n_H^4$

\item $\sum_{l \neq m} |(n_l)_z |^2  |(n_m)_z |^2  = (3/2) n_H^4$

\end{itemize}
With $\int n_x^4 = (3/8)n_H^4$ and $\int n_z^4 = n_0^4 + 6n_0^2
n_H^2 +(9/4) n_H^4$, the $W$-energy becomes

\ba \mathrm{SkX2}: ~ W (n_0^4 \!+\! 6 n_0^2 n_H^2 \!+\! 3 n_H^4).
\ea
SkX3 state gives
\ba  4n_0\int \Phi_z^3 =  6 n_0 (n_{\v k_1})_z(n_{\v k_2})_z (n_{\v
k_3})_z \nn = 6 n_0 n_H^3 \cos(\theta_1 \!+\! \theta_2 \!+\!
\theta_3) .\ea
For $\int \Phi_x^4=\int \Phi_y^4 = (81/64) n_H^4$ one works out

\begin{itemize}

\item $\sum_\alpha | (n_\alpha)_x |^4  = (27/64) n_H^4$

\item $\sum_{l \neq m} | (n_l)_x |^2  | (n_m)_x |^2 = (27/32) n_H^4$

\end{itemize}
For $\int \Phi_z^4 = (45/8)n_H^4$ one works out

\begin{itemize}

\item $\sum_\alpha  | (n_\alpha)_z |^4 = (9/8) n_H^4$

\item $\sum_{l \neq m}
 | (n_l)_z |^2  |(n_m)_z |^2  = (9/2) n_H^4$

\end{itemize}
The $W$-energy is $\int n_x^4 = (81/64)n_H^4$, $\int n_z^4 = n_0^4 +
6 n_0 n_H^3 \cos(\theta_1\!+\!\theta_2\!+\!\theta_3) \!+\! 9 n_0^2
n_H^2 \!+\! (45/8)n_H^4$, and

\ba \mathrm{SkX3}: ~ W\!\Bigl(\!n_0^4 \!+\! 6 n_0 n_H^3 \cos(\theta_1
\!+\! \theta_2 \!+\! \theta_3)\nn
\!+\! 9n_0^2 n_H^2 \!+\! {261 \over 32}n_H^4 \!\Bigr). \ea
The phase angle dependence $\sim \cos(\theta_1 \!+\! \theta_2 \!+\!
\theta_3)$ arises in the $W$-energy as well as in the $U$-energy for
SkX3.

SC state has two types of terms contributing to $\int \Phi_x^4=\int
\Phi_y^4=\int \Phi_z^4=(9/4)n_H^4$:

\begin{itemize}

\item $\sum_\alpha  | (n_\alpha)_z |^4 = (3/4) n_H^4$

\item $\sum_{l \neq m}
 | (n_l)_z |^2 | (n_m)_z |^2 = (3/2) n_H^4$

\end{itemize}
The $W$-energy is $\int n_x^4 = (9/4)n_H^4$, $\int n_z^4 = n_0^4 + 6
n_0^2 n_H^2 + (9/4)n_H^4$, and

\ba \mathrm{SC}:~ W\Bigl(n_0^4 \!+\! 6 n_0^2 n_H^2 \!+\! {27 \over
4}n_H^4\Bigr) .\ea

FCC state has contributions to $\int \Phi_x^4=\int \Phi_y^4=\int
\Phi_z^4 =(14/3)n_H^4 $ coming from

\begin{itemize}

\item $\sum_\alpha  | (n_\alpha)_x |^4 = (2/3)n_H^4$

\item $\sum_{l \neq m} | (n_l)_x |^2  | (n_m)_x |^2 = 4 n_H^4$

\end{itemize}
The $W$-energy is $\int n_x^4 = (14/3)n_H^4$, $\int n_z^4 = n_0^4
+8n_0^2n_H^2 + (14/3)n_H^4$ and

\ba \mathrm{FCC}: ~  W\Bigl(n_0^4 + 8n_0^2 n_H^2 +14n_H^4 \Bigr).\ea

BCC state has
\ba && 4n_0 \int \Phi_z^3 =\sum_{\v q_1 \!+\!\v q_2\!+\!\v q_3 = \v
0} n_0 (n_{\v q_1})_z(n_{\v q_2})_z (n_{\v q_3})_z \nn
&& = 3n_0n_H^3 \! \Bigl[\cos(\theta_1 \!+\! \theta_4 \!-\! \theta_5)
\!+\! \cos(\theta_3 \!+\! \theta_6  \!-\! \theta_1) \nn
&& ~~~~~\!-\! \cos(\theta_2 \!+\! \theta_5 \!-\! \theta_3) \!-\!
\cos(\theta_2 \!+\! \theta_4 \!+\! \theta_6)\Bigr] .\ea
Calculation of $\int \Phi_x^4 =\int \Phi_y^4=\int \Phi_z^4
=(87/8)n_H^4$ can be obtained from

\begin{itemize}

\item $\sum_\alpha  | (n_\alpha)_x |^4 = (9/8)n_H^4$

\item $\sum_{l \neq m}
 | (n_l)_x |^2  | (n_m)_x |^2 = (78/8) n_H^4$

\end{itemize}
The $W$-energy is $\int n_x^4 = (87/8)n_H^4$ and

\ba && \mathrm{BCC}: ~  W\Bigl(n_0^4 + 12n_0^2 n_H^2 +{261 \over
8}n_H^4 \nn
&&~~ \!+\! 3n_0n_H^3 \bigl[\cos(\theta_1 \!+\! \theta_4 \!-\!
\theta_5) \!+\! \cos(\theta_3 \!+\! \theta_6 \!-\! \theta_1) \nn
&&~~~~ \!-\! \cos(\theta_2 \!-\! \theta_5 \!-\! \theta_3) \!-\!
\cos(\theta_2 \!+\! \theta_4 \!+\! \theta_6)\bigr]\Bigr).\ea

\section{Minimizing the phase angles}
The three phase angles of SkX3 enter in the energy as

\ba ( 9U + 6W ) n_0 n_H^3 \cos(\theta_1 + \theta_2 + \theta_3) .\ea
It is minimized by choosing $\theta_1 + \theta_2 +\theta_3 = \pi$.

The six phase angles appearing in the BCC energy can be grouped as
\ba && X \Bigl[  \cos (\theta_1 \!+\!\theta_4 \!-\!\theta_5 \!+\!
\alpha_1 ) \!+\! \cos (\theta_3 \!+\!\theta_6 \!-\!\theta_1 \!+\!
\alpha_2 ) \nn
&& + \cos (\theta_2 \!+\!\theta_5 \!-\!\theta_3 \!+\! \alpha_3 )\!
+\! \cos (\theta_2 \!+\!\theta_4 \!+\!\theta_6 \!+\! \alpha_4 )
\Bigr] \nn \ea
with $X=\sqrt{27U^2 +30UW + 9W^2}$ and $\alpha_i$ that can be worked
out. The minimum energy configuration is achieved by taking all the
arguments of the cosines equal to $\pi$. The energy of BCC coming
from the $U$ and $W$ terms after minimizing the phase angles are

\ba \mathrm{BCC}:&&   U\Bigl(n_0^4 \!+\!14 n_0^2 n_H^2 \!+\! 54 n_H^4
\Bigr)\nn
&&+ W\Bigl(n_0^4 \!+\! 12n_0^2 n_H^2 \!+\!{261 \over 8}n_H^4 \Bigr)
\nn
&& - 4\sqrt{27U^2 \!+\! 30UW \!+\! 9W^2} n_0 n_H^3.\ea
No other candidate states besides SkX3 and BCC have the phase angles
entering in the energy.

\end{document}